%% file: EXO-11-019_temp.tex
\pdfoutput=1

\documentclass[11pt,twoside,a4paper,cmspaper,final,collab]{cms-tdr}
\def\svnVersion{167116}\def\cmsCernNo{2012-157}\def\cmsCernDate{\today}\def\cmsMessage{Submitted to Physics Letters B}

\begin{document}\cmsNoteHeader{EXO-11-019}

\hyphenation{had-ron-i-za-tion}
\hyphenation{cal-or-i-me-ter}
\hyphenation{de-vices}
\RCS$Revision: 128293 $
\RCS$HeadURL: svn+ssh://svn.cern.ch/reps/tdr2/papers/EXO-11-019/trunk/EXO-11-019.tex $
\RCS$Id: EXO-11-019.tex 128293 2012-06-08 19:50:04Z alverson $
\newlength\cmsFigWidthFive
\ifthenelse{\boolean{cms@external}}{\setlength\cmsFigWidthFive{\columnwidth}}{\setlength\cmsFigWidthFive{0.5\textwidth}}
\newlength\cmsFigWidthSix
\ifthenelse{\boolean{cms@external}}{\setlength\cmsFigWidthSix{\columnwidth}}{\setlength\cmsFigWidthSix{0.6\textwidth}}
\newlength\cmsFigWidthSeven
\ifthenelse{\boolean{cms@external}}{\setlength\cmsFigWidthSeven{\columnwidth}}{\setlength\cmsFigWidthSeven{0.7\textwidth}}
\ifthenelse{\boolean{cms@external}}{\providecommand{\cmsLeft}{top}}{\providecommand{\cmsLeft}{left}}
\ifthenelse{\boolean{cms@external}}{\providecommand{\cmsRight}{bottom}}{\providecommand{\cmsRight}{right}}
\cmsNoteHeader{EXO-11-019} 

\newcommand{\numtrigs}{$10^{7}$}
\newcommand{\anaLumiabstract}{5\fbinv}
\newcommand{\anaLumi}{5.28 $\pm$ 0.12\fbinv}
\newcommand{\anaLumiee}{4.98 $\pm$ 0.11\fbinv}
\newcommand{\energyresolutioneb}{1.0\%}
\newcommand{\energyresolutionee}{2.3\%}
\newcommand{\control}{$120 < M_{\ell\ell} < 200\GeV$}
\newcommand{\searchRegion}{$600\GeV$}
\newcommand{\obseventscontrol}{XXX}
\newcommand{\expectedeventscontrolnonDY}{XXX}
\newcommand{\expectedeventscontrolDY}{XXX}

\newcommand{\FakeContEleA}{$381 \pm 153$}
\newcommand{\FakeContEleB}{$127 \pm 51$}

\newcommand{\efficelecEB}{87.3\% $\pm$ 1.2\%}
\newcommand{\efficelecEE}{82.0\% $\pm$ 1.3\%}
\newcommand{\efficSFEB}{1.007 $\pm$ 0.011}
\newcommand{\efficSFEE}{1.016 $\pm$ 0.010}

\newcommand{\efficMbarrel}{$95.0\%\pm0.3\%$}
\newcommand{\efficMendcap}{$90.0\%\pm0.4\%$}
\newcommand{\efficSFmuB}{0.972 $\pm$ 0.003}
\newcommand{\efficSFmuE}{0.966 $\pm$ 0.004}
\newcommand{\efficMuTrig}{$92.4\% \pm 0.3\%$}

\newcommand{\etcutee}{35}
\newcommand{\ptcutMu}{35}

\newcommand{\nevtDielePeak}{3}
\newcommand{\nevtDimuPeak}{4}
\newcommand{\nevtPeak}{7}

\newcommand{\MassComb} {$ M_{\ell\ell} = xxx \pm yyy \pm zzz ~{\mathrm{TeV}}$}
\newcommand{\CrossSecComb} {$ \sigma \times B (pp\rightarrow X\rightarrow\ell^+\ell^-) = xx \pm xx \pm xx ~{\mathrm{pb}}$}

\newcommand{\SSMwidth}{30}
\newcommand{\Esixwidth}{6}
\newcommand{\Gonewidth}{14}
\newcommand{\Gtwowidt}{3.5}
\newcommand{\Stwidth}{0.06}

\newcommand{\limitmumuZssm}{2150}
\newcommand{\limitmumuZpsi}{1820}
\newcommand{\limitmumuGlow}{1630}
\newcommand{\limitmumuGhigh}{1990}
\newcommand{\limiteeZssm}{2120}
\newcommand{\limiteeZpsi}{1810}
\newcommand{\limiteeGlow}{1640}
\newcommand{\limiteeGhigh}{1960}
\newcommand{\limitZssm}{2330}
\newcommand{\limitZpsi}{2000}
\newcommand{\limitGlow}{1810}
\newcommand{\limitGhigh}{2140}
\newcommand{\limitStsix}{890}
\newcommand{\limitStthree}{445}
\newcommand{\limitStfour}{540}

\newcommand{\fix}[1]{{\bf <<< #1 !!! }}%
\providecommand{\GKK}{\ensuremath{\mathrm{G}_\text{KK}}}
\providecommand{\tq}{\ensuremath{\mathrm{t}}}
\providecommand{\ZPSSM}{\ensuremath{\cPZpr_\text{SSM}}}
\providecommand{\ZPPSI}{\ensuremath{\cPZpr_\psi}}
\providecommand{\ZPSt}{\ensuremath{\cPZpr_\text{St}}}
\providecommand{\ZPSteps}{0.06}
\providecommand{\re}{\ensuremath{\mathrm{e}}} 

\title{Search for narrow resonances in dilepton mass spectra in pp collisions at $\sqrt{s}$ = 7\TeV} 

\date{\today}

\abstract{
A search for narrow, high-mass resonances decaying to electron or
muon pairs has been performed using pp collision data collected at
$\sqrt{s} = 7\TeV$ by the CMS experiment in 2011. The data
sample corresponds to an integrated luminosity of approximately
\anaLumiabstract.
 The event yields observed in the signal regions are consistent with
predictions of the standard model backgrounds, and upper limits on the
cross section times branching fraction for a resonance decaying to
dileptons are extracted from a shape analysis of the dilepton
invariant mass distribution. The resulting mass limits at 95\% confidence level  are
\limitZssm\GeV for the $\zp$ in the Sequential Standard Model, \limitZpsi\GeV for the
superstring-inspired $\ZPPSI$ resonance,
\limitStsix\ (\limitStfour)\GeV for the Stueckelberg extension $\ZPSt$ with the mass parameter
  $\epsilon=\ZPSteps$ (0.04), and \limitGhigh\ (\limitGlow)\GeV for
  Kaluza--Klein gravitons with the coupling parameter
  $k/\overline{M}_\text{Pl}$ of 0.10 (0.05). These limits are the most
  stringent to date.
}

\hypersetup{%
pdfauthor={CMS Collaboration},%
pdftitle={Search for narrow resonances in dilepton mass spectra in pp collisions at sqrt(s) = 7 TeV},%
pdfsubject={CMS},%
pdfkeywords={CMS, dileptons, resonances}}

\maketitle 

\section{Introduction}

This Letter describes the results of a search for narrow resonances in
the dilepton mass spectra using data collected by the Compact Muon
Solenoid (CMS) detector at the Large Hadron Collider (LHC)~\cite{lhc}
at CERN during 2011.

Numerous models describing possible physics beyond the standard model
(SM) predict the existence of narrow resonances at the TeV mass
scale.  The results of the search reported here are interpreted in
the context of several such models.  The narrow resonances predicted
in these models include the Sequential Standard Model $\ZPSSM$ with
standard model  couplings~\cite{Altar:1989}, the $\ZPPSI$ expected
in grand unified theories~\cite{Leike:1998wr}, the $\ZPSt$ produced in
the Stueckelberg extension to the standard
model~\cite{Kors2004,Feldman2006a}, and the Kaluza--Klein graviton
($\GKK$) excitations arising in the Randall--Sundrum (RS) model of
extra dimensions~\cite{Randall:1999vf, Randall:1999ee}.  For a
resonance mass of 1\TeV, the widths of the $\ZPSSM$, $\ZPPSI$,
$\ZPSt$, and $\GKK$ are \SSMwidth, \Esixwidth, 0.06, and \Gtwowidt\
(\Gonewidth)\GeV, where the $\GKK$ coupling parameter
$k/\overline{M}_\text{Pl}$ is taken to be 0.05 (0.1).
In the case of the $\ZPSt$ this width is evaluated at
a value of $\ZPSteps$ for the $\epsilon$
parameter. This parameter represents the ratio of the mass parameters of the gauge
bosons before they mix to become the $\cPZ$ and \ZPSt.
The maximum allowed value of $\epsilon$ is approximately 0.06,
as determined by the precision measurements of the $\cPZ$ parameters.
If $\epsilon$ is equal to 0, there is
no coupling to the Stueckelberg extension and the standard model is
recovered.

Results of searches for narrow $\zp \rightarrow \ell^+ \ell^-$ and
$\GKK \rightarrow \ell^+ \ell^-$ resonances
have previously been reported by the ATLAS~\cite{ATLAS-dilep-2011} and CMS~\cite{original_paper}
collaborations, based on about 1\fbinv and  40\pbinv of data,
respectively. The D0 and CDF experiments have published results with
over 5\fbinv of integrated luminosity in $\Pp\Pap$ collisions,
 at a centre-of-mass energy
of 1.96\TeV~\cite{D0_RS,D0_Zp,CDF_RS,CDF_Zp,CDF_SSM,CDF_RSele}.
Indirect constraints have been placed on
the mass of virtual $\cPZpr$ bosons by LEP-II
experiments~\cite{delphi,aleph,opal,l3}, using the cross
sections and angular distributions of dilepton and hadronic final
states in $\Pep\Pem$ collisions.

The results presented in this Letter are obtained from an analysis of
pp collision data at $\sqrt{s} = 7\TeV$
 corresponding to an integrated luminosity of
\anaLumi\ for the muon channel and \anaLumiee\ for
the electron channel~\cite{LumiRef2011}.
The analysis procedure reported here follows methods used in the earlier analysis
of a smaller data set, described in Ref.~\cite{original_paper}.
The search for resonances is based on a shape analysis of the dilepton
mass spectra, to be robust against uncertainties in the absolute background level.
In the absence of a signal, limits are set on the ratio
$R_{\sigma}$ of the production cross section times branching fraction
for high-mass resonances to that for the Z boson. In this approach, many
experimental and theoretical uncertainties common to both measurements
cancel.  Using theoretical cross sections and including K factors and parton
distribution functions (PDFs), lower mass limits are calculated for several models.

\section{The CMS detector}\label{sec:CMS}
The central feature of the CMS detector is a superconducting solenoid
providing an axial magnetic field of 3.8\unit{T} and enclosing the all-silicon inner
tracker, the crystal electromagnetic calorimeter (ECAL), and the
brass-scintillator hadronic calorimeter (HCAL).  The inner tracker is
composed of a pixel detector and a silicon strip tracker, and measures
charged-particle trajectories in the pseudorapidity range
$|\eta|<2.5$\,\footnote{{A right-handed coordinate system is used in
CMS, with the origin at the nominal collision point, the $x$ axis
pointing to the center of the LHC ring, the $y$ axis pointing up
(perpendicular to the LHC plane), and the $z$ axis along the
anticlockwise-beam direction. The pseudorapidity $\eta$ is defined as
$\eta = -\ln \tan (\theta/2)$, where $\cos \theta = p_z/p$. The
azimuthal angle $\phi$ is the angle relative to the positive $x$ axis
measured in the $x$-$y$ plane.}}.  The finely segmented ECAL consists
of nearly 76\,000 lead-tungstate crystals which provide coverage in
pseudorapidity up to $|\eta| = 3.0$.  The muon system covers the
pseudorapidity region $|\eta|<2.4$ and consists of up to four stations
of gas-ionization muon detectors installed outside the solenoid and
sandwiched between steel layers serving both as hadron absorbers and
as a return yoke for the magnetic field.  A detailed description of
the CMS detector can be found
elsewhere~\cite{JINST}.

The CMS experiment uses a two-level trigger system. The \Lone
Trigger, composed of custom hardware processors, selects events
of interest using information from the calorimeters and muon
detectors~\cite{L1TDR}.  The High-Level Trigger (HLT) is software-based
and further decreases the event collection rate by using the full event
information, including that from the inner tracker~\cite{HLTTDR}.

\section{Electron and muon selection \label{sec:lepton}}

The events used in the dimuon channel analysis were collected using a
single-muon trigger with a transverse momentum ($\pt$) threshold of 40\GeV.  In order to keep
the trigger rate at an acceptable level, the acceptance of this trigger
was restricted to the pseudorapidity range of $|\eta| < 2.1$.
The muon candidates' tracks are formed in the trigger by combining standalone tracks
reconstructed separately in the muon chambers and in the inner tracker.

The trigger used to select dielectron events
requires the presence of two clusters in the
ECAL, each with transverse energy $\ET > 33\GeV$ and each matched
to hits in the pixel detector. The trigger also requires the
absence of significant energy deposits in the hadron calorimeter cells
directly behind these two ECAL clusters.

Electrons and muons are reconstructed using standard CMS algorithms,
described in more detail
in~\cite{MUO-10-002-PAS,EWK-10-002-PAS,original_paper}. Clusters in the
ECAL are matched to reconstructed tracks to form electron candidates.
These candidates must be within the barrel or endcap acceptance
regions, with pseudorapidities of $|\eta|<1.442$ and
$1.560<|\eta|<2.5$, respectively. Electron candidates must have $\ET >
\etcutee$\GeV if they are within the barrel region and $\ET >
40$\GeV if they are within the endcap regions.  As in the muon
trigger, muon tracks are reconstructed separately  in both the
muon system and the inner tracker~\cite{MUO-10-002-PAS} and then
matched and fitted simultaneously to form ``global muons''.
Each of the muon candidates must have $\pt > 45\GeV$; the candidate
must also have a transverse impact
parameter with respect to the centre of the luminous region of less than 0.2\cm, at
least one hit in the pixel detector, hits in at least nine silicon
tracker layers, and matched segments in two or more muon stations.
The muon candidates are required to come from the same vertex
by performing a common-vertex fit and requiring the vertex
$\chi^2$ to be below 10.

To suppress the misidentification of jets as electrons, the sum
of the $\pt$ of all other tracks in a cone of $\Delta R =
\sqrt{(\Delta\eta)^2 + (\Delta\phi)^2} < 0.3$ around the electron
candidate's track must be less than 5\GeV, and the sum of the $\ET$
of calorimeter energy deposits in the same cone must be less than
approximately 3\% of the candidate's $\ET$.
For the calculation of the $\pt$ sum, tracks must pass
within 0.2\cm (in the $z$ direction) of the primary vertex with which the electron candidates are associated.
With respect to the earlier analysis~\cite{original_paper} and as a
consequence of the increase in the number of interaction per bunch crossing,
the longitudinal segmentation of the HCAL in the endcaps is
no longer used to identify electrons.  To suppress both jets and
non-prompt muon sources of misidentification for muons, the sum of
the $\pt$s of all other tracks within a cone of $\Delta R < 0.3$
about the muon candidate's track must be less than 10\% of the
candidate's $\pt$.

The data sample collected in 2011 has been used to improve the alignment
of the muon chambers. The dimuon mass resolution,
$\sigma( m_{\Pgm\Pgm})/m_{\Pgm\Pgm}$,
is
6.5\% at masses around 1\TeV, rising to 12\% at 2\TeV.  The fractional
dielectron mass resolution, $\Delta m_{\Pe\Pe}/m_{\Pe\Pe}$, is approximately
constant above 500\GeV.  When both electrons
are detected in the barrel, this mass resolution is 1.1\%,
and when one of the electrons is in the barrel and the other is in the endcaps it is 2.3\%

While a knowledge of the overall triggering and identification
efficiencies is required to set limits on specific models, only the
energy dependence of these efficiencies needs to be evaluated for the
measurement of $R_{\sigma}$.  The triggering and particle
identification efficiencies in the energy range up to about 150\GeV
were measured from data using the ``tag-and-probe"
method~\cite{EWK-10-002-PAS,MUO-10-002-PAS}.   Monte Carlo
(MC) simulations were used to
evaluate the evolution of efficiencies beyond this energy.
For dielectron events, the combined efficiency of the first level and
high level triggers is larger than
99\% and requires no corrections. The data were used to measure
the electron identification efficiency at the Z resonance, and the
ratio of this efficiency to that found in the simulation is
used to scale the efficiencies at high energies. This ratio is within
1\% of unity.  The efficiency at $\pt=100\GeV$ is ($86 \pm 2$)\% in the
barrel and ($84 \pm 2$)\% in the endcaps.  For simulated events, the
variation in this efficiency is less than 2\% for $\pt$ above 100\GeV.
A similar procedure
was used to extract scale factors for the muon trigger efficiency and
for muon identification.  The trigger efficiency was found to be $(91.3 \pm 0.1)\%$,
and the muon identification
efficiencies were found to be ($96.3 \pm 0.2$)\% and ($94.2 \pm 0.2)$\%
for the barrel and endcaps, respectively, where the uncertainties given are statistical only.
These uncertainties remain uniform over the $\pt$
region that was probed by applying the tag-and-probe method to
the available data sample.

\section{Event samples and event selection}

Simulated event samples for the signal and background processes were
variously generated with \PYTHIA, \MADGRAPH and \POWHEG.
The \MADGRAPH~\cite{MADGRAPH} matrix-element generator was used for
$\ttbar$, single top and $|PW + \text{jets}$ samples and the
\POWHEG~V1.1 framework~\cite{Alioli:2008gx, Nason:2004rx,
Frixione:2007vw}  for Drell--Yan to electrons and muons and single top samples.
Both of these were interfaced with the \PYTHIA~V6.424 (using the Z2
tune)~\cite{Sjostrand:2006za,PythiaZ2Tune} parton-shower
generator. All other processes were generated using \PYTHIA.
The {CTEQ6L1}~\cite{Pumplin:2002vw}
parton distribution function (PDF) set was used for all samples except the Drell--Yan where
the CT10~\cite{Lai:2010} set was used.  The response of the detector
was simulated in detail using \GEANTfour~\cite{GEANT4}. These samples
were further processed through the trigger emulation and event
reconstruction chain of the CMS experiment.

For both the dimuon and dielectron final states, two isolated,
same-flavour leptons that pass the lepton identification criteria
described in Section~\ref{sec:lepton} were required.  The two lepton charges
were required to be of opposite sign in the case of dimuons (for which
a charge misassignment implies a large momentum measurement error),
but not in the case of dielectrons (where charge assignment is
decoupled from the ECAL-based energy measurement).  An opposite-charge
requirement for dielectrons would lead to a loss of signal efficiency
of a few percent and hence was not applied.

The electron event selection requires the presence of at least one electron candidate in the
ECAL barrel because events with both electrons in the endcaps have a
lower signal-to-background ratio as a result of a higher rate of jets being misidentified as electrons.
For both channels, each event was
required to have a reconstructed vertex with at least four associated
tracks, located less than 2\cm from the centre of the detector in the
direction transverse to the beam and within 24\cm in the direction
along the beam. This requirement supresses cosmic ray background.
Additional suppression of cosmic ray muons was obtained by
requiring the three-dimensional opening angle between the two muons to
be smaller than $\pi - 0.02$ radians.

\section {Backgrounds}

The most prominent SM process that contributes to the dimuon and
dielectron invariant mass spectra is Drell--Yan production
($\cPZ{/}\gamma^*$), either directly or via $\tau\tau$; there are also contributions from $\ttbar$,
$\tq\PW$, and diboson processes. In addition,
jets may be misidentified as leptons and contribute to the dilepton
invariant mass spectra through multijet and vector boson plus jets final
states. The contamination from diphotons misidentified as dielectrons, as well
possible contributions from $\cPqb\cPaqb$ and $\cPqc\cPaqc$ events,
have been established to be negligible.

In the final dilepton spectra, the background component from standard model
processes is found by fitting an appropriate function to the data. To
find an appropriate functional form, trial variants were fitted to
distributions obtained from MC simulations. The studies of the
background components described below were performed in order to verify
that the assumed background composition is correct and are not used directly
to estimate the magnitude of the background.

\subsection{\texorpdfstring{\cPZ/$\gamma^*$}{Z/gamma*} backgrounds}

The shape of the dilepton invariant mass spectrum from
Drell--Yan production was obtained using a MC simulation based on the \POWHEG event
generator.  The simulated invariant mass spectrum was normalized to the
data using the number of events in the mass interval of 60--120\GeV.
The shape of this spectrum can be modified by higher-order
corrections and by variations in PDFs. An uncertainty due to these sources
is assigned to the extrapolation of the background shape, from masses
where no non-standard model contribution is expected, to higher
masses.  The procedure used is described in Ref.~\cite{original_paper}. The
uncertainty in the predicted number of events normalized to those expected in
the $\cPZ$ peak ranges from approximately 5\% at a mass of 400\GeV to
20\% at a mass of 2\TeV.

\subsection{Other backgrounds with prompt lepton pairs
\label{sec:e-mu}  }

\begin{figure}[htbp]
\centering
\includegraphics[width=\cmsFigWidthSix]{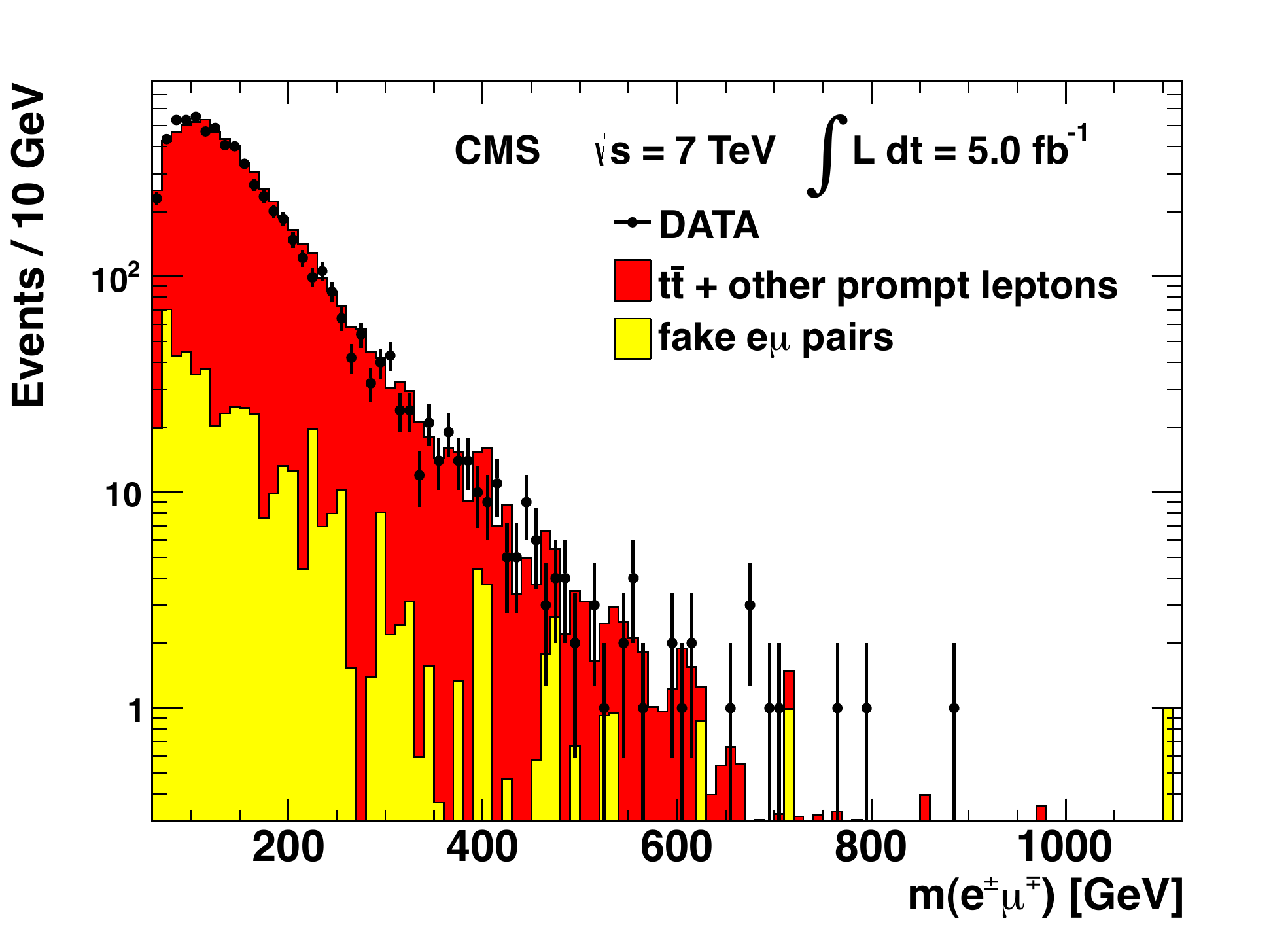}
\caption{\label{fig:muonselectrons}
The observed opposite-sign $\Pe^\pm\Pgm^\mp$ dilepton invariant mass spectrum
(data points). 
The filled red histogram shows the contribution to the spectrum from
$\ttbar$ and other sources of prompt leptons ($\tq\PW$, diboson
production, $\cPZ\to\Pgt\Pgt$), as  derived from simulations.
The background where at least one of the reconstructed objects is not a real lepton
is shown in yellow and estimated from the data using the same-sign $\Pe^\pm\Pgm^\pm$ spectrum.
}
\end{figure}

Pairs of prompt leptons can arise from {$\ttbar$}, $\tq\PW$, and
diboson production. These processes are lepton flavour symmetric, enabling
the use of an $\Pe\Pgm$ spectrum to assess the contribution of these
processes to the same-flavour dilepton spectra. The invariant mass
spectrum found using a trigger that requires the presence of both a
photon (or electron) and a muon is shown in
Fig.~\ref{fig:muonselectrons}.  Using a single-muon trigger, a very
similar spectrum is found.  The leptons in this figure are required to
have opposite signs. The components of the background arising from
real leptons are estimated from MC simulations.  The background
contribution arising from jets being misidentified and reconstructed as leptons
is derived from data by using same-sign $\Pe\Pgm$ spectrum.  The
observed number of $\Pe\Pgm$ events with any sign combination allowed
is 3863 (1175) in the mass region above 120 (200)\GeV.  Using MC
simulation, and the data for the contribution where at least one jet
has been misreconstructed as a lepton, the expected number of events
above 120 (200)\GeV is $4081 \pm 406$ ($1305 \pm 123$).
The overall uncertainty in these numbers is dominated by the theoretical uncertainty of
15\% on the $\ttbar$ production cross
section~\cite{Campbell:2010ff,Kleiss:1988xr}. Note that these numbers are not used to
determine the final mass limits.

\subsection{Events with misidentified or non-prompt leptons}

Candidate prompt leptons can be misreconstructed from tracks and energy deposits that
have not originated from a lepton.
The misidentification of jets as
leptons, the principal source of such backgrounds, is more likely to
occur for electrons than for muons. The muon background from this source
was determined as in Ref.~\cite{original_paper} and verified to be negligible
(less than 0.05 events above 600\GeV).

Both jets and photons can be misidentified as electrons. Potential
sources of such backgrounds are $\PW\to \Pe\cPgn + \text{jet}$, $\gamma +
\text{jet}$ events and multijet events. A single electromagnetic-cluster trigger
collected a sample of events used to determine the fraction
of jets passing the electromagnetic trigger criteria that are misreconstructed as
electrons.  To suppress the contribution from $\cPZ$ decays, events in
this sample are required to have no more than one reconstructed
electron passing less stringent than standard selection criteria.  Contamination from
genuine electrons in $\PW$+jet events and from converted photons in
$\gamma+\text{jet}$ events may affect the misidentification rate measurement. The
contributions from these sources were estimated using MC simulations and
subtracted from the data, to perform this measurement.
The mass spectrum due to events with at least one misidentified electron
was found by summing the multijet spectrum estimated from the data and
the $\PW\to \Pe\cPgn + \text{jet}$ and $\gamma +\text{jet}$ contributions estimated
using MC simulations.  The multijet spectrum was found by using an
event sample passing the trigger used to collect signal events and applying
the probability that both candidates are misidentified as electrons.
The magnitude of this total contribution is
illustrated in Fig.\ \ref{fig:spectra}.
The estimated background contribution to the dielectron
mass spectrum due to misidentified jets is \FakeContEleA\
(\FakeContEleB)
for $m_{\Pe\Pe} > 120$ (200)\GeV.

\subsection{Cosmic ray muon backgrounds}

The $\Pgmp\Pgmm$ data sample is susceptible to contamination from
traversing cosmic ray muons, which may be misreconstructed as a pair
of oppositely charged, high-momentum muons.  Cosmic ray events are
removed from the data sample using selection criteria mentioned above,
which eliminate events with two muons having collinear tracks and
events with muons that have large impact parameters relative to the
collision vertex.  For the dimuon mass region $m_{\Pgm\Pgm} >
200$\GeV, the residual mean expected background was estimated using
two event samples.
Events in one sample were selected without imposing the requirement on
the dimuon opening angle and in the other sample the requirements on
muon impact parameter and on the existence of a good quality primary
vertex were not applied.
The efficiencies of the remaining cuts were
estimated using these samples and treated as uncorrelated in order to
determine the final total efficiency. This background was found to be less than
0.2 events.

\section{Dilepton invariant mass spectra}

\begin{figure}[htbp]
\centering
\includegraphics[width=0.45\textwidth]{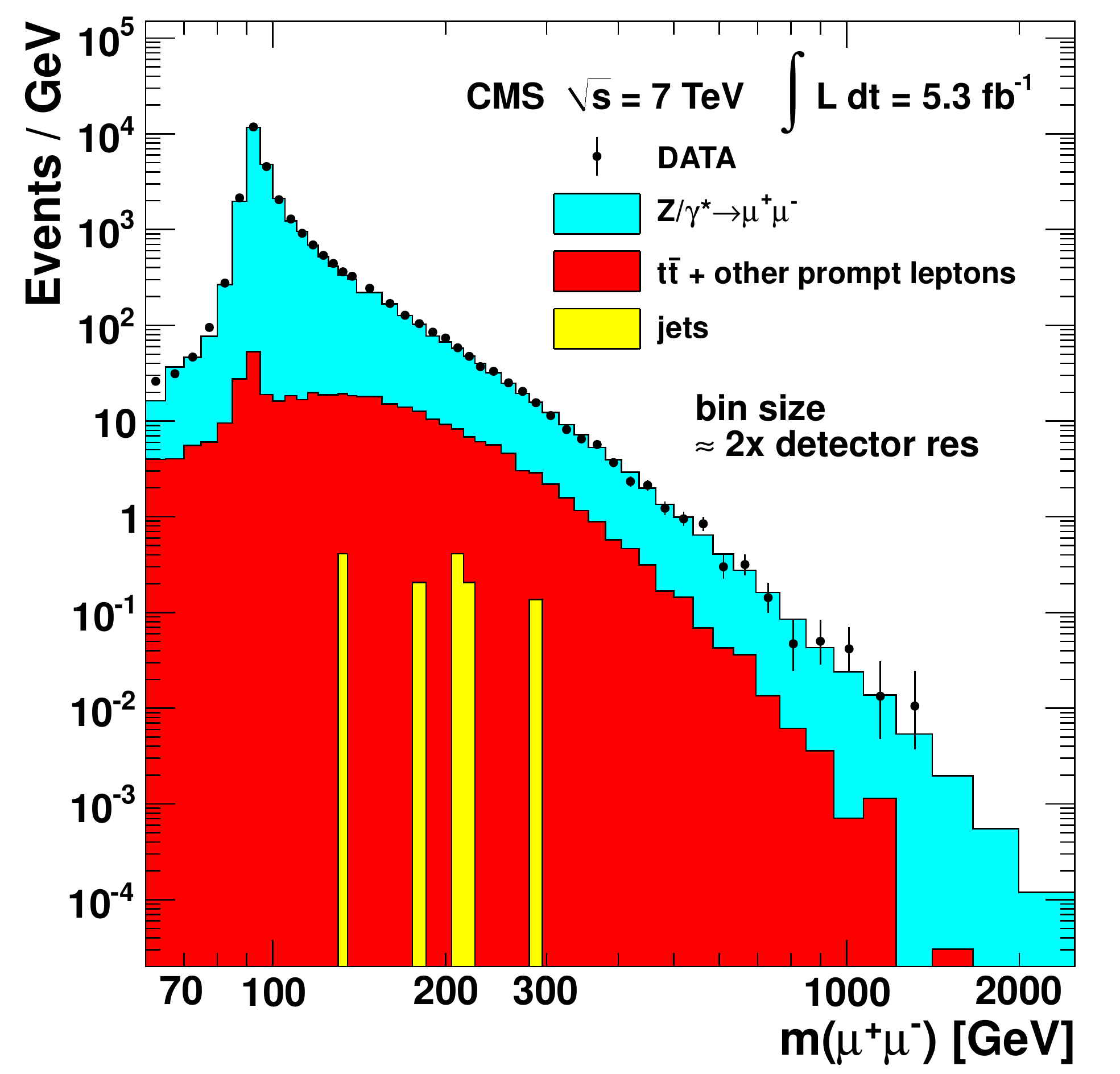}
\includegraphics[width=0.45\textwidth]{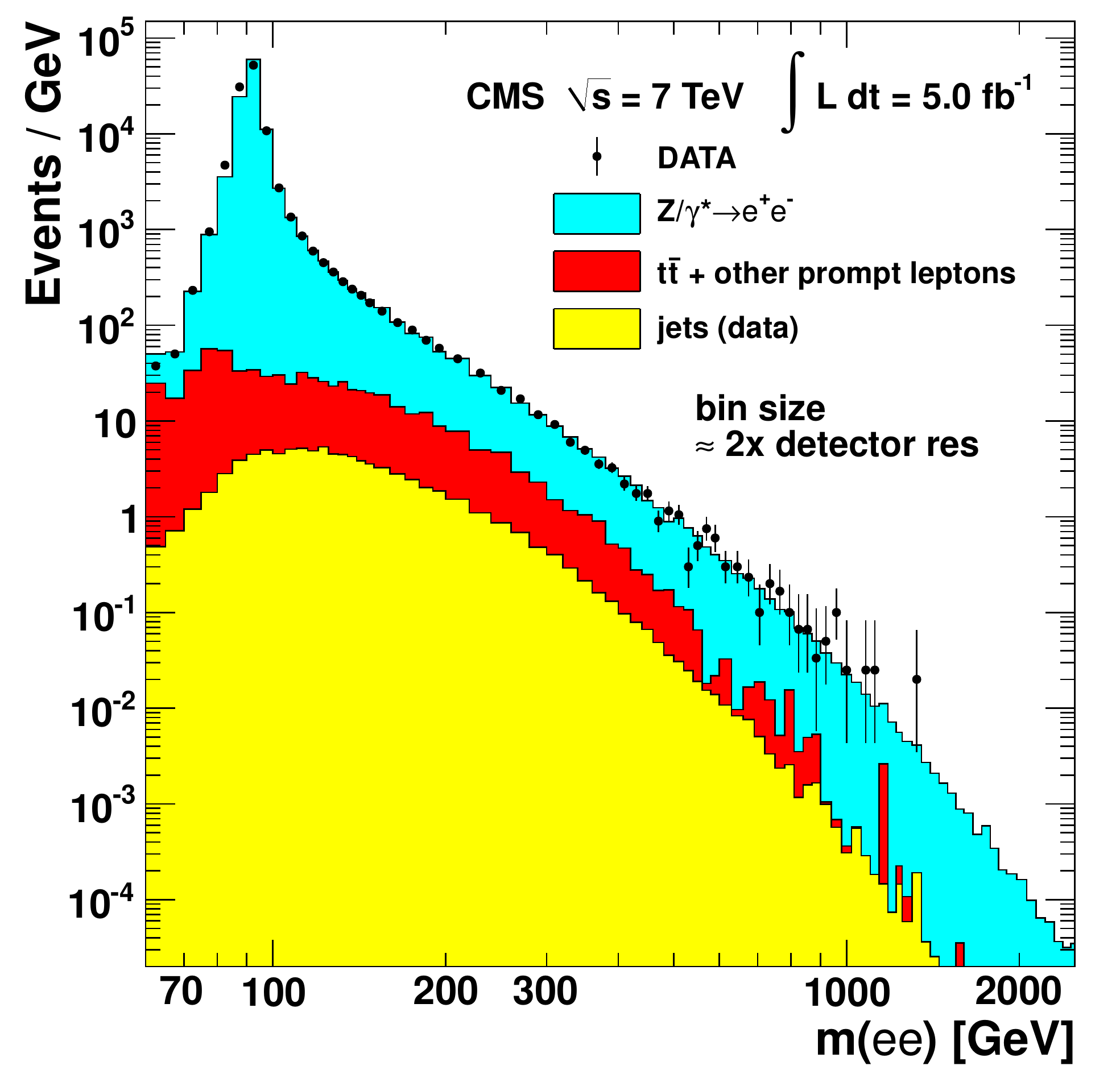}
\caption{\label{fig:spectra}
The invariant mass spectrum of $\Pgmp\Pgmm$ (\cmsLeft) and $\Pe\Pe$ (\cmsRight) events. The points with
error bars represent data. The uncertainties in the data points are statistical only.
The
histograms represent the expectations from
SM processes: $\cPZ{/}\gamma^*$, $\ttbar$ and other sources of
prompt leptons ($\tq\PW$, diboson production, $\cPZ\to\Pgt\Pgt$), and
the multijet backgrounds. Multijet backgrounds contain at least one
jet that has been misreconstructed as a lepton.
}
\end{figure}

\begin{figure}[htbp]
\centering
\includegraphics[width=0.45\textwidth]{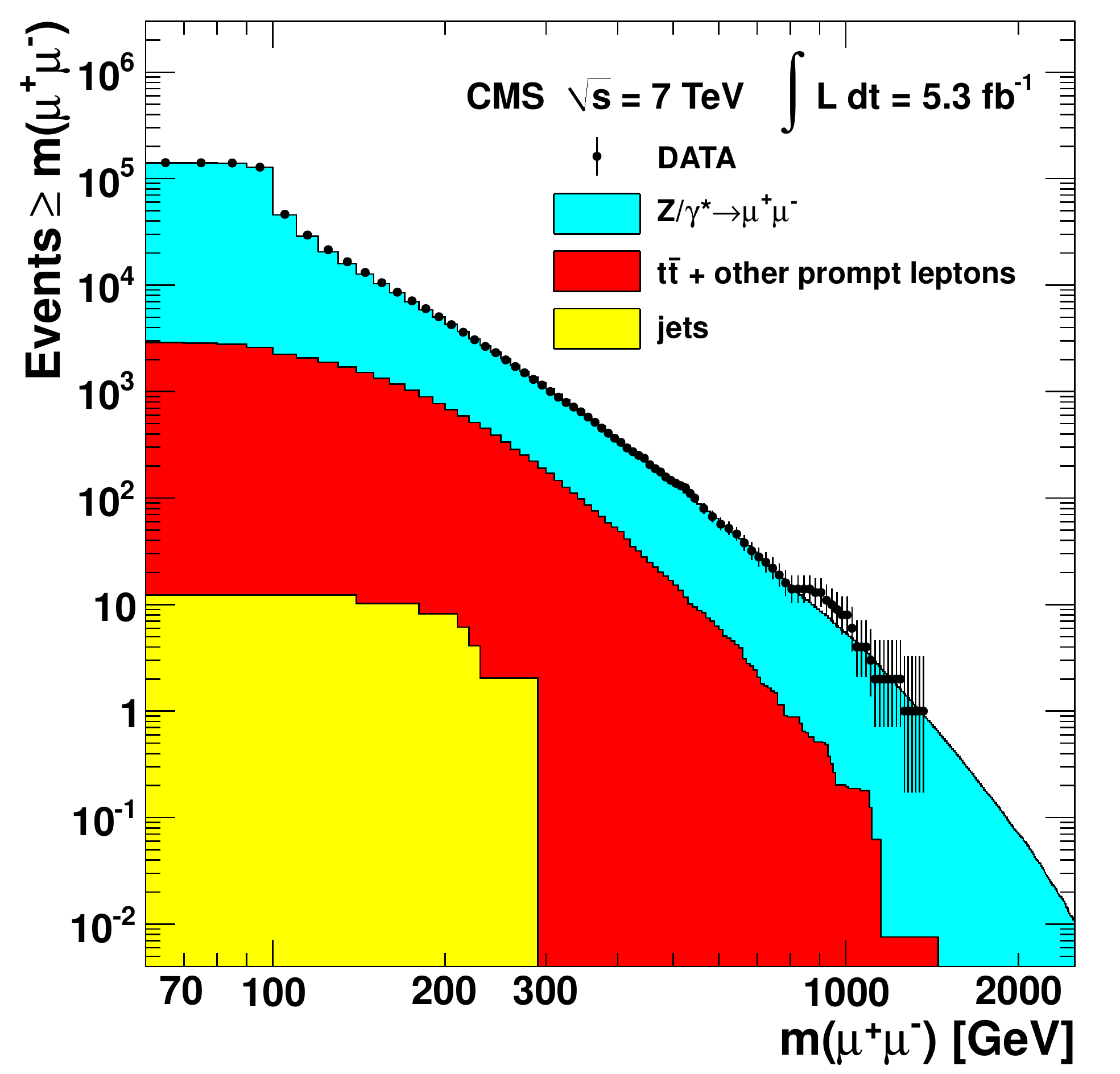}
\includegraphics[width=0.45\textwidth]{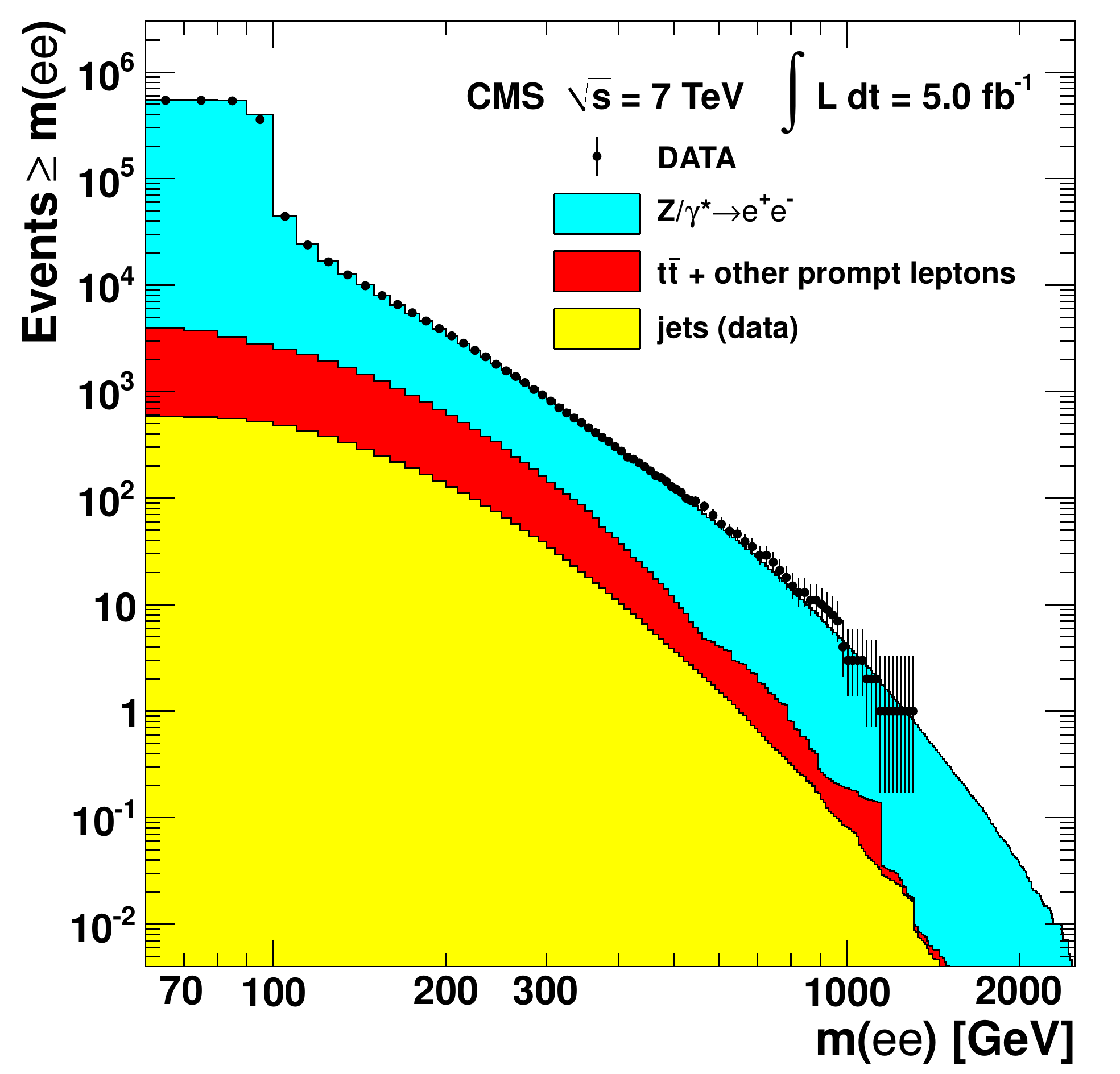}
\caption{\label{fig:cum_spectra}
The cumulative distribution of the invariant
mass spectrum of $\Pgmp\Pgmm$ (\cmsLeft) and $\Pe\Pe$ (\cmsRight) events.  The
points with error bars represent data; the histograms
represent the expectations from SM processes.  }
\end{figure}

Figure~\ref{fig:spectra} shows a comparison of data and expected
backgrounds in both dimuon (\cmsLeft) and dielectron (\cmsRight) mass spectra.
The illustrated ``jets" contribution includes events where at least one jet
has been misreconstructed as a lepton.
The component from events where two jets are misreconstructed as electrons
was obtained from data.  Contributions from
$\PW\to \Pe\cPgn + \text{jet}$ and $\gamma + \text{jet}$ events were estimated from MC simulations,
as were all other backgrounds illustrated.
The relative fractions of backgrounds derived from simulation
are determined using theoretical cross sections.
Overall, these backgrounds are normalized to
the data using the ratio of the number of observed to expected events
within a window of 60--120\GeV, which includes the Z resonance peak.
Figure~\ref{fig:cum_spectra} shows the corresponding cumulative
distributions of the spectra for the dimuon (\cmsLeft) and dielectron
(\cmsRight) samples.  The expected yields in the control region
(120--200\GeV) and in the high invariant mass region (${>}200\GeV$) are
listed in Table~\ref{tab:event_yield}.
The observed data agree with the expectations. (It should be noted that
such agreement is not critical to the shaped-based analysis discussed
below.)

\begin{table*}[htbp]
\centering
\topcaption{
The number of dilepton events with invariant mass in the control
region $120 < m_{\ell\ell} < 200$\GeV and in the search region $m_{\ell\ell} > 200$\GeV.
The total background is the sum of the events for the SM processes listed.
The yields from simulation are relatively normalized using the expected cross sections,
and overall the simulation is normalized to the data using the number of events in the mass window 60--120\GeV.
Uncertainties include both statistical and systematic components
added in quadrature. }
\label{tab:event_yield}
\begin{tabular}{l|c|c|c|c}
\hline\hline
Source & \multicolumn{4}{c}{Number of events} \\
       & \multicolumn{2}{c|}{Dimuon sample }
       & \multicolumn{2}{c}{Dielectron sample} \\
 &  ($120-200$)\GeV  & $ >$200\GeV
 &  ($120-200$)\GeV  & $ >$200\GeV  \\ \hline
Data                       &  17240             &  4250              & 13207             & 3335      \\
Total background           & $16272 \pm 739$     & $4266 \pm 185$   & $13286 \pm 625$    & $3209 \pm 276$ \\ \hline
$\cPZ{/}\gamma^*$          & $15055 \pm 726$     &  $3591 \pm 170$   & $11945 \pm 597$    & $2615\pm 262$   \\
$\ttbar$ + other prompt leptons & $1213 \pm 145$   & $667 \pm 80$    & $1087\pm 163$      & $467 \pm 70$ \\

Sources including at least & $4 \pm 3$           & $8 \pm 4$             & $254 \pm 102   $   & $127\pm 51$  \\

one jet misreconstructed  & & & & \\
as a lepton & & & & \\

\hline\hline
\end{tabular}
\end{table*}

The cross check procedures and the event scrutiny described in
Ref.~\cite{original_paper} were performed for all events
with an invariant mass above 800\GeV. No anomalies
were found.

\section{Statistical analysis and results}

The observed invariant mass spectra agree with expectations based on
standard model processes. Limits are set on the possible
contributions from narrow heavy resonances as follows.

\begin{figure}[htbp]
\centering
\includegraphics[width=\cmsFigWidthSeven]{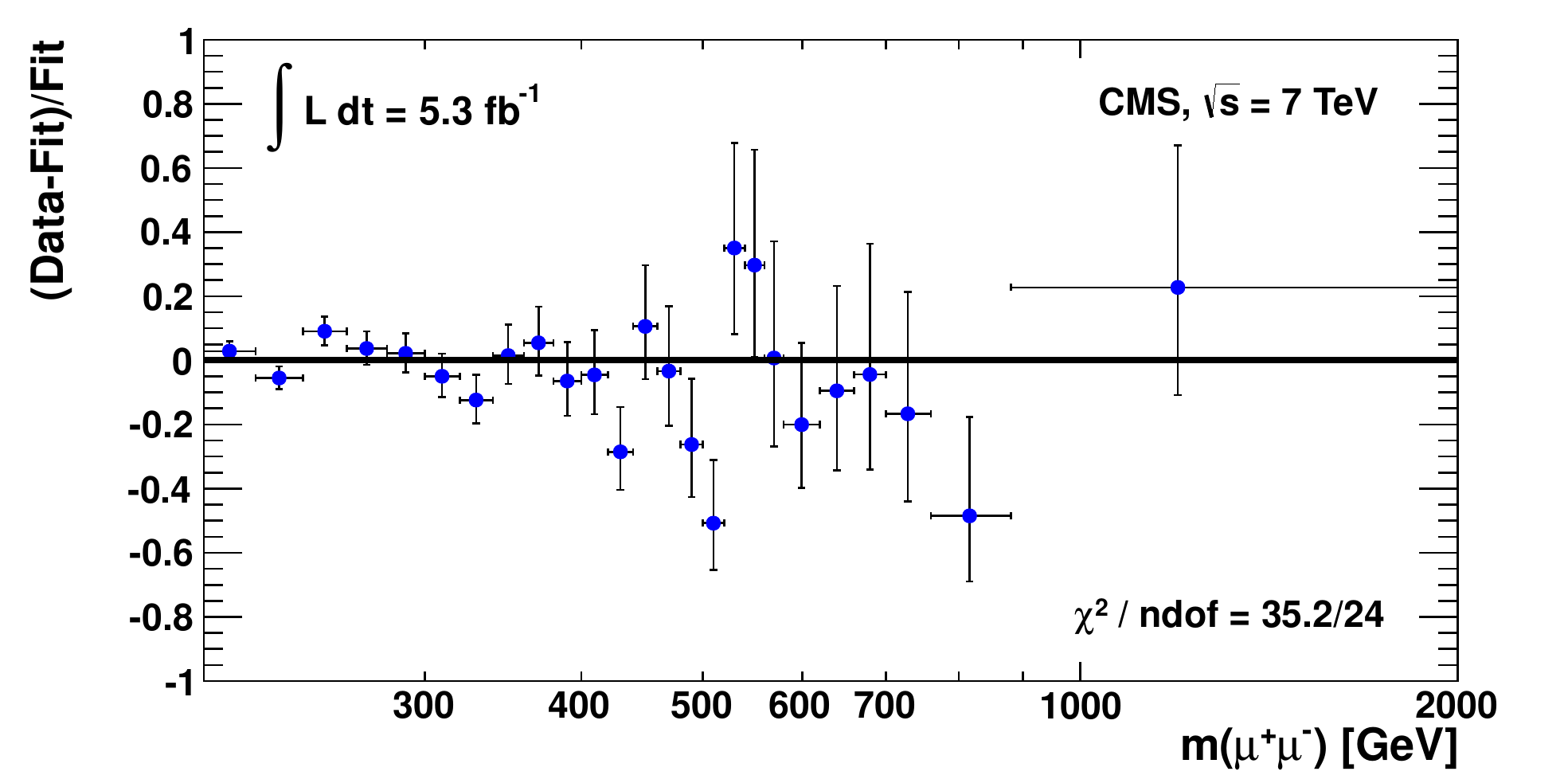}
\includegraphics[width=\cmsFigWidthSeven]{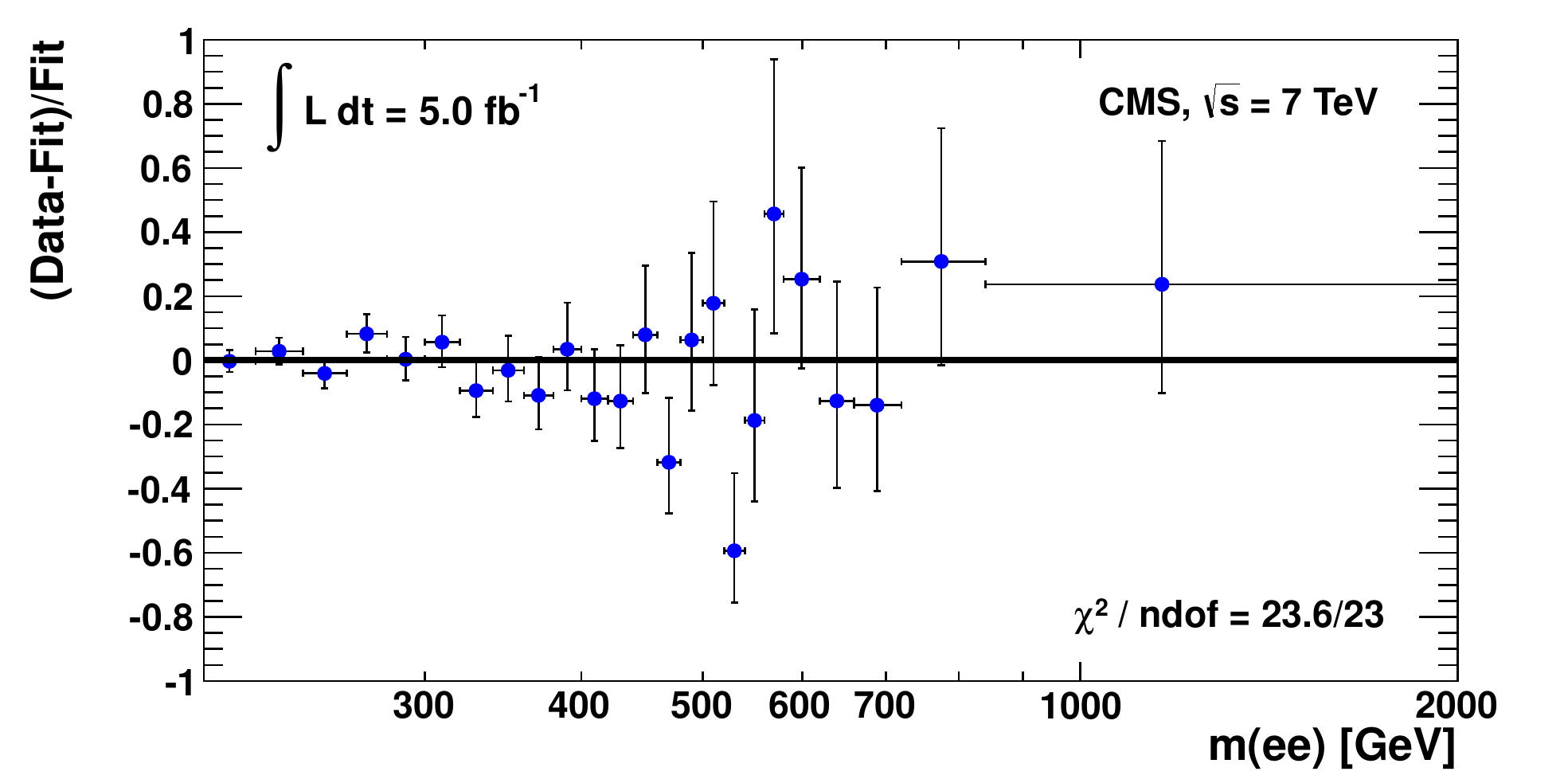}
\caption{\label{fig:dataMinusFit}
The relative difference between the data and the fitted parametrization of the simulated
background, where the latter is normalized to the data, is shown
in a variety of mass bins for the
muon (top) and electron (bottom) channels. The binning was chosen so that
there is a minimum prediction of 10 events in each bin and a minimum
bin size of 20\GeV was required. The horizontal error bars simply represent the bin
width and should not be intepreted as an uncertainty. }
\end{figure}

The procedure followed to set 95\% confidence level (CL) limits is
identical to that described in Ref.~\cite{original_paper}.  An
extended unbinned likelihood function is used based on a signal shape,
parametrized by a Breit-Wigner function convolved with a Gaussian
resolution function, and a background function with approximately
exponential behaviour. The functional form used for the background is
$m^{-\kappa}\re^{-\alpha m}$, where the shape parameters $\kappa$ and
$\alpha$ were determined from a fit to a simulated background mass
spectrum. The agreement between this fit and the observed data over a
range of mass bins is shown in Fig.~\ref{fig:dataMinusFit}. The
width of the Breit--Wigner is conservatively taken to be that of the
$\cPZpr$ boson in the Sequential Standard Model, which has the largest
width of the resonances considered, namely 3.1\%. The highest mass
limits are insensitive to this width.

An upper limit on the ratio $R_{\sigma}$ of the cross section times
branching fraction of a $\cPZpr$ boson relative to that for a Z boson
was found using the Bayesian technique described
in Ref.~\cite{original_paper}.  The dominant uncertainty in this analysis is that
in $R_\epsilon$, the ratio of selection efficiency times detector
acceptance for $\cPZpr$ decay to that for $\cPZ$ decay. This uncertainty is 8\% for
the dielectron channel and 3\% for the dimuon channel. These values
reflect the current understanding of the detector acceptance and
the reconstruction efficiency turn-on at low mass (including PDF
uncertainties on the acceptance),
as well as their values at high mass. The largest contribution to the
dielectron uncertainty comes from the uncertainty in the electron
reconstruction efficiency at high mass as, unlike for muons where
cosmic rays provide a high momentum sample, there is no available data
sample of high energy electrons with high purity.  For the dielectron
channel, the mass scale uncertainty is 1\%, derived from a study of
the variation of the energy scale with time and a linearity study
using different methods to calculate the ECAL cluster
energies~\cite{original_paper}.  For the dimuon channel, the mass
scale uncertainty for this analysis is set to zero. A sensitivity
study showed negligible change in the limits up to the maximum
detector misalignment consistent with alignment studies, a shift
corresponding to a several percent change in the momentum scale.
The effects of the uncertainties in the PDF and the higher order
corrections ~\cite{Melnikov:2006kv}
on the shape of the background distribution, and hence on the fitting
function used for the background, were examined. No further systematic
uncertainties were found to be required to accommodate these effects
on the background shape. In the
electron channel the background from jets misidentified as electrons is very
small, and the uncertainty in this background has a negligible
effect on the limit determination.  The acceptance for $\GKK$ (spin 2)
is higher than for $\cPZpr$ (spin 1) by less than 8\% over the mass
range 0.75--2.0\TeV. This difference in acceptance was conservatively
neglected when calculating the corresponding limits.

\begin{figure}[htbp]
\centering
\includegraphics[width=\cmsFigWidthSix]{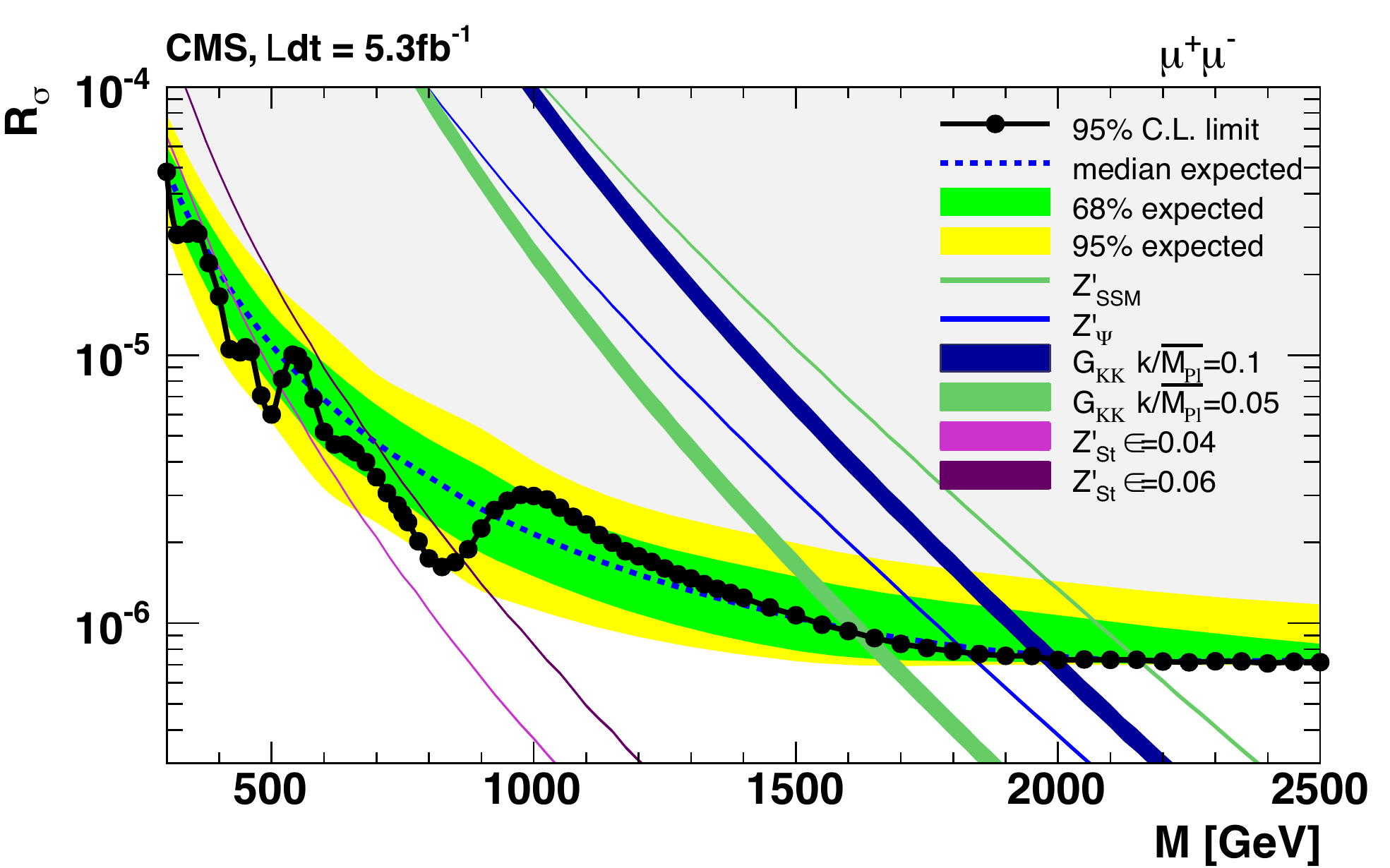}
\includegraphics[width=\cmsFigWidthSix]{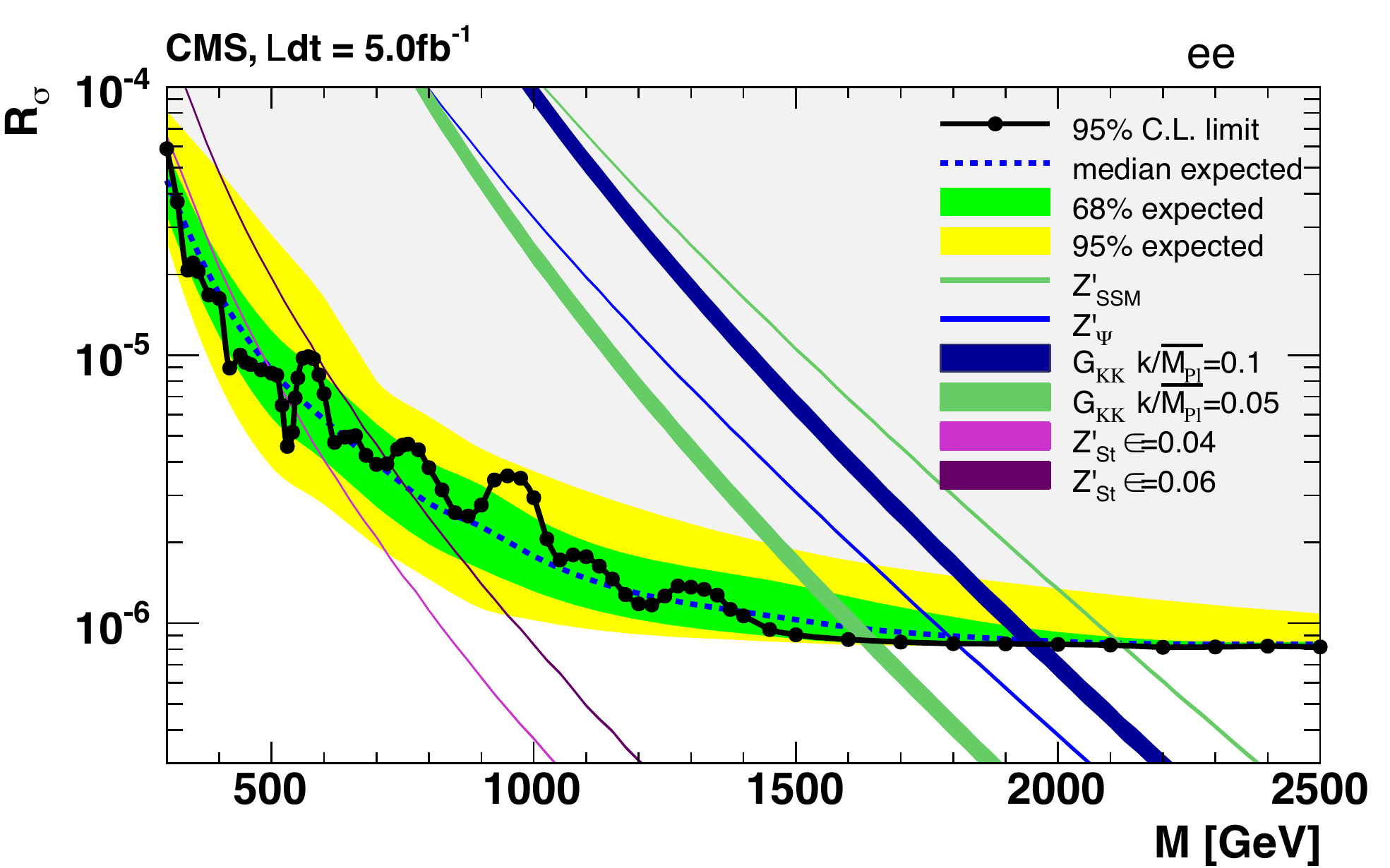}
\includegraphics[width=\cmsFigWidthSix]{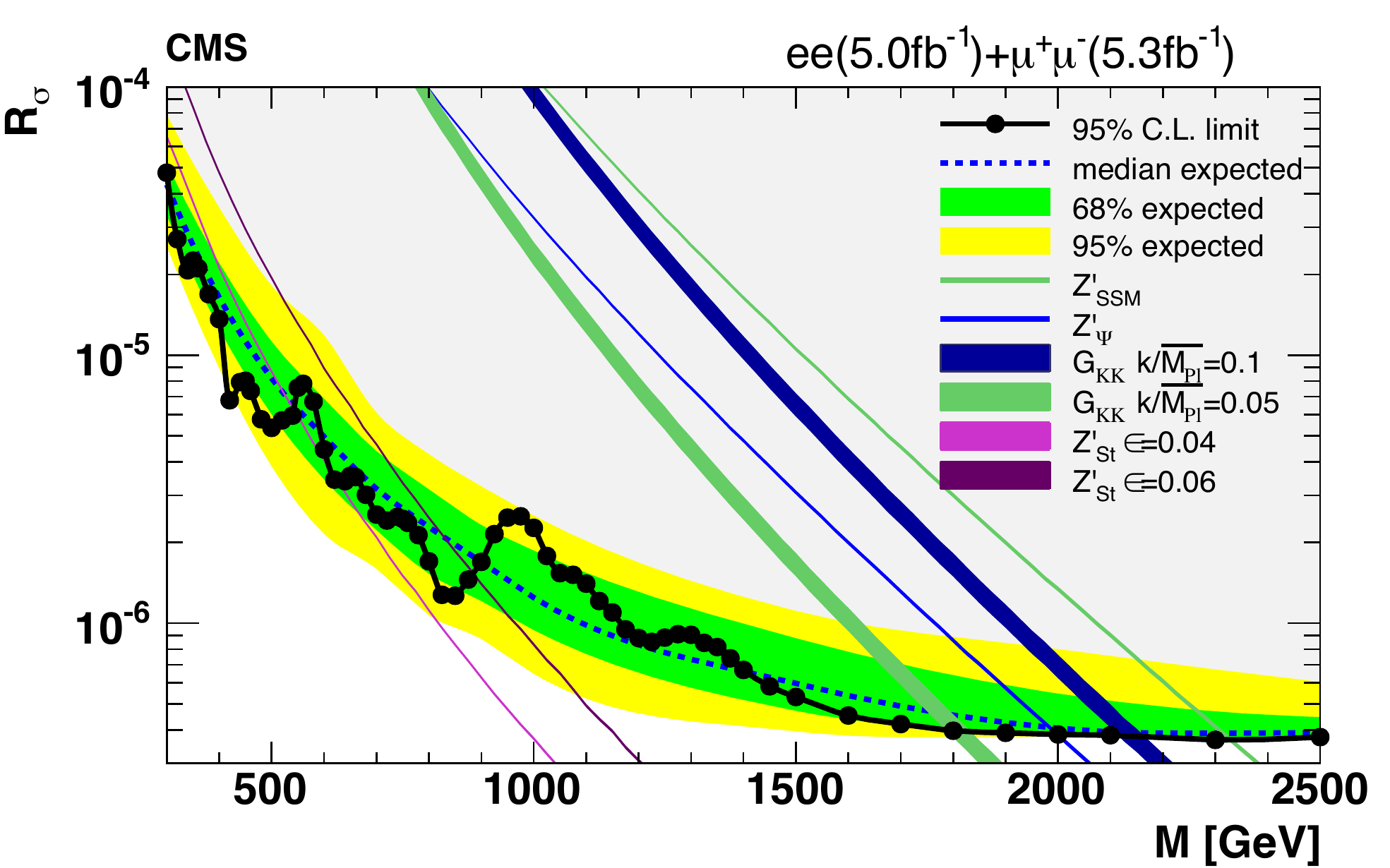}
\caption{\label{fig:limits} Upper limits as a function of the resonance
  mass $M$ on the production ratio $R_{\sigma}$ of cross section
  times branching fraction into lepton pairs for $\ZPSSM$, $\ZPPSI$, $\ZPSt$,
  and $\GKK$ production to the same quantity for Z bosons.
The limits are shown from
  (top) the $\Pgmp\Pgmm$ final state, (middle) the $\Pe\Pe$ final
  state and (bottom) the combined dilepton result.  Shaded green and
  yellow bands correspond to the $68\%$ and $95\%$ quantiles for the
  expected limits.  The predicted cross section ratios are shown as
  bands, with widths indicating the theoretical uncertainties.
  The differences in the widths reflect the different uncertainties in the
  K factors used.}
\end{figure}

In Fig.~\ref{fig:limits}, the predicted cross section times branching fraction ratios for
$\ZPSSM$, $\ZPPSI$, and $\ZPSt$ production are shown  together with those
for $\GKK$ production, with the dimensionless graviton coupling to SM
fields $k/\overline{M}_\mathrm{Pl}=0.05$ and $0.1$.  The leading-order
cross section predictions for $\ZPSSM$, $\ZPPSI$, and\ $\ZPSt$ from \PYTHIA using
{CTEQ6.1} PDFs are corrected for a mass dependent K factor
obtained using \textsc{zwprodp}~\cite{Accomando:2010fz,Hamberg:1990np,
vanNeerven:1991gh}, to account for the next-to-next-to-leading order
(NNLO) QCD contributions.  For
the RS graviton model, a constant next-to-leading order K factor of 1.6 is
used~\cite{Mathews:2005bw}. The uncertainties due to
factorization and renormalization scales
and PDFs are indicated as a band.
The calculated $\cPZpr$ and $\GKK$ cross sections include generated
dileptons with masses only within $\pm$40\% of the nominal resonance
mass~\cite{original_paper,Accomando:2010fz}.
The NNLO prediction for the $\cPZ{/}\gamma^*$
production cross section in the mass window of 60 to 120\GeV is
$0.97\pm0.04\unit{nb}$~\cite{Melnikov:2006kv}.

The uncertainties described above are propagated into a comparison of the experimental
limits with the predicted cross section times branching fraction ratios
($R_{\sigma}$) to obtain 95\%~CL lower limits on $\cPZpr$ masses in various models.
No uncertainties on cross sections for the various theoretical models
are included when determining the limits.
As a result of the dimuon
analysis, the $\ZPSSM$ can be excluded below \limitmumuZssm\GeV, the
$\ZPPSI$ below \limitmumuZpsi\GeV, and the RS $\GKK$ below
\limitmumuGhigh\ (\limitmumuGlow)\GeV for couplings of 0.10 (0.05). For
the dielectron analysis, the production of $\ZPSSM$ and $\ZPPSI$
bosons is excluded for masses below \limiteeZssm\ and
\limiteeZpsi\GeV, respectively. The corresponding lower limits on the
mass for RS $\GKK$ with couplings of 0.10 (0.05) are \limiteeGhigh\
(\limiteeGlow)\GeV.

\begin{figure}[htbp]
\centering
\includegraphics[width=\cmsFigWidthSix]{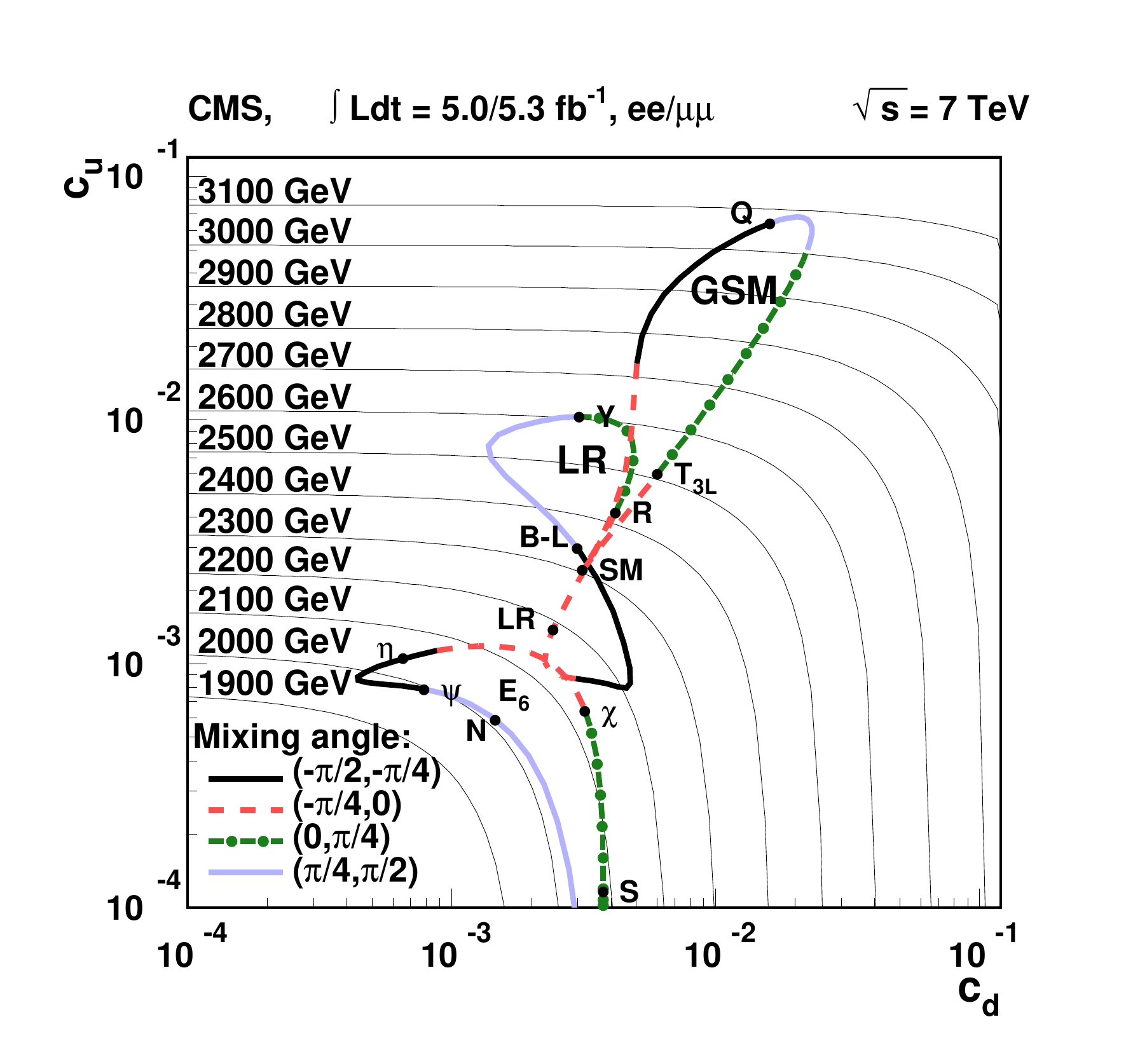}
\caption{\label{fig:CuCd}
The 95\% CL upper limits on the \cPZpr\ cross sections for given masses
are equivalent to excluded regions in the $(c_\cPqd,c_\cPqu)$ plane which are
bounded by the thin black lines in the figure.
They are compared with the predicted values of $(c_\cPqd,c_\cPqu)$ in three
classes of models.
The colours on curves
correspond to different mixing angles of the generators defined in each model.
For any point on a model curve, the mass limit
corresponding to that value of $(c_\cPqd,c_\cPqu)$ is given by the
intersecting experimental contour.
}
\end{figure}

The combined limit, obtained by using the product of the likelihoods
for the individual channels, is shown in Fig.~\ref{fig:limits} (bottom plot). The
signal cross section is constrained to be the same in the two channels
and lepton universality is assumed.  The 95\% CL lower limits on the
mass of a $\cPZpr$ resonance are \limitZssm\GeV for $\ZPSSM$,
\limitZpsi\GeV for $\ZPPSI$, and\ \limitStsix\ (\limitStfour)\GeV for $\ZPSt$ with
$\epsilon=\ZPSteps$ (0.04). The RS Kaluza--Klein gravitons are excluded below
\limitGhigh\ (\limitGlow)\GeV, for couplings of 0.10 (0.05).  The
observed limits are more restrictive than those previously obtained
via similar direct searches by the Tevatron
experiments~\cite{D0_RS,D0_Zp,CDF_RS,CDF_Zp,CDF_RS_2011,CDF_Zp_2011}
and indirect searches by LEP-II
experiments~\cite{delphi,aleph,opal,l3}, as well as those obtained by
ATLAS~\cite{ATLAS-dilep-2011} and CMS~\cite{original_paper} using
smaller data samples. The results are also presented in the
$(c_\cPqd,c_\cPqu)$ plane in Fig.~\ref{fig:CuCd} \cite{Carena:2004xs, original_paper}.
The parameters $c_\cPqd$
and $c_\cPqu$ contain all the information about the model-dependent
couplings of the \cPZpr\ to fermions in the annihilations of charge $-1/3$ and
2/3 quarks, respectively.  The cross-section limits at any
particular mass are contours in the $(c_\cPqd,c_\cPqu)$ plane. The model
classes are described in \cite{original_paper,Accomando:2010fz}.

The largest deviation from SM expectations at high masses is around
1\TeV, in both spectra. The statistical significance of the
observations is expressed in terms of $Z$-values, which are the
effective numbers of Gaussian standard deviations in a one-sided test.  For the
dimuon sample, the maximum excess occurs at 1005\GeV, with local
$Z=1.2$, while for the dielectron sample, the maximum excess occurs at
960\GeV, with local $Z=1.7$.  In the combination of the two channels,
the maximum excess is found at 965\GeV, with local $Z=2.1$.  The
probability of an enhancement at least as large as the one found
occurring anywhere between 600 and 2500\GeV in the observed sample
size corresponds to $Z=-0.7$ for the dimuon sample and $Z=0.3$ for the
dielectron sample. For the combined data sample, the corresponding
probability in a joint peak search is equivalent to $Z=0.4$.

\section{Summary}

The CMS Collaboration has searched for narrow resonances in dimuon and
dielectron invariant mass spectra using
pp collision data collected at $\sqrt{s}$ = 7\TeV,
corresponding to an integrated luminosity of \anaLumi for dimuons and
\anaLumiee for dielectrons.
The spectra are consistent with expectations from the
standard model and upper limits have been set on the cross section
times branching fraction for $\cPZpr$ into lepton pairs relative to
standard model $\cPZ$ boson production and decay.
Lower mass limits have been set at 95\%~CL:
a Sequential Standard Model $\cPZpr$ can be excluded below
\limitZssm\GeV, the superstring-inspired $\ZPPSI$ below
\limitZpsi\GeV, the $\ZPSt$ with an $\epsilon$ parameter of \ZPSteps\ (0.04)
below \limitStsix\ (\limitStfour)\GeV, and Randall--Sundrum Kaluza--Klein gravitons below \limitGhigh\
(\limitGlow)\GeV for couplings of 0.10 (0.05).  The
constraints on these cross sections and masses are the most stringent
to date.

\newpage
\section*{Acknowledgments}
\hyphenation{Bundes-ministerium Forschungs-gemeinschaft  Forschungs-zentren}
We congratulate our colleagues in the CERN accelerator departments for
the excellent performance of the LHC machine. We thank the technical
and administrative staff at CERN and other CMS institutes, and
acknowledge support from: FMSR (Austria); FNRS and FWO (Belgium);
CNPq, CAPES, FAPERJ, and FAPESP (Brazil); MES (Bulgaria); CERN; CAS,
MoST, and NSFC (China); COLCIENCIAS (Colombia); MSES (Croatia); RPF
(Cyprus); MoER, SF0690030s09 and ERDF (Estonia); Academy of Finland,
MEC, and HIP (Finland); CEA and CNRS/IN2P3 (France); BMBF, DFG, and
HGF (Germany); GSRT (Greece); OTKA and NKTH (Hungary); DAE and DST
(India); IPM (Iran); SFI (Ireland); INFN (Italy); NRF and WCU (Korea);
LAS (Lithuania); CINVESTAV, CONACYT, SEP, and UASLP-FAI (Mexico); MSI
(New Zealand); PAEC (Pakistan); MSHE and NSC (Poland); FCT (Portugal);
JINR (Armenia, Belarus, Georgia, Ukraine, Uzbekistan); MON, RosAtom,
RAS and RFBR (Russia); MSTD (Serbia); SEIDI and CPAN (Spain); Swiss
Funding Agencies (Switzerland); NSC (Taipei); TUBITAK and TAEK
(Turkey); STFC (United Kingdom); DOE and NSF (USA).

Individuals have received support from the Marie-Curie programme and
the European Research Council (European Union); the Leventis
Foundation; the A. P. Sloan Foundation; the Alexander von Humboldt
Foundation; the Belgian Federal Science Policy Office; the Fonds pour
la Formation \`a la Recherche dans l'Industrie et dans l'Agriculture
(FRIA-Belgium); the Agentschap voor Innovatie door Wetenschap en
Technologie (IWT-Belgium); the Council of Science and Industrial
Research, India; the Compagnia di San Paolo (Torino); and the HOMING
PLUS programme of Foundation for Polish Science, cofinanced from
European Union, Regional Development Fund.

\bibliography{auto_generated}   

\cleardoublepage \appendix\section{The CMS Collaboration \label{app:collab}}\begin{sloppypar}\hyphenpenalty=5000\widowpenalty=500\clubpenalty=5000\input{EXO-11-019-authorlist.tex}\end{sloppypar}
\end{document}

%% file: EXO-11-019-authorlist.tex
\textbf{Yerevan Physics Institute,  Yerevan,  Armenia}\\*[0pt]
S.~Chatrchyan, V.~Khachatryan, A.M.~Sirunyan, A.~Tumasyan
\vskip\cmsinstskip
\textbf{Institut f\"{u}r Hochenergiephysik der OeAW,  Wien,  Austria}\\*[0pt]
W.~Adam, T.~Bergauer, M.~Dragicevic, J.~Er\"{o}, C.~Fabjan, M.~Friedl, R.~Fr\"{u}hwirth, V.M.~Ghete, J.~Hammer, N.~H\"{o}rmann, J.~Hrubec, M.~Jeitler, W.~Kiesenhofer, V.~Kn\"{u}nz, M.~Krammer, D.~Liko, I.~Mikulec, M.~Pernicka$^{\textrm{\dag}}$, B.~Rahbaran, C.~Rohringer, H.~Rohringer, R.~Sch\"{o}fbeck, J.~Strauss, A.~Taurok, P.~Wagner, W.~Waltenberger, G.~Walzel, E.~Widl, C.-E.~Wulz
% ...
\vskip\cmsinstskip
\dag:~Deceased\\
% ...
